\documentclass[runningheads]{llncs}
\usepackage[T1]{fontenc}
\usepackage{graphicx}
\usepackage[most]{tcolorbox}
\usepackage{graphicx,array}
\usepackage{xcolor}
\usepackage{fancyvrb} 
\usepackage{listings}
\usepackage{multirow}
\usepackage{tabularx}

\begin{document}
\title{ModelForge: Using GenAI to Improve the Development of Security Protocols}
%
%
%
\newcommand\sudip[1]{\textcolor{red}{Sudip: #1}}
\newcommand\ivan[1]{\textcolor{orange}{Ivan: #1}}
\newcommand\martin[1]{\textcolor{purple}{Martin: #1}}

\newcommand{\LLMOne}{Codestral}
\newcommand{\LLMTwo}{Claude 3.5}
\newcommand{\LLMThree}{GPT-4o}
\newcommand{\LLMFour}{LLama 3.1}
\newcommand{\LLMFive}{ModelForge}
\author{Martin Duclos\inst{1}\orcidID{0000-0003-2182-6496} \and
Ivan A. Fernandez\inst{1}\orcidID{0009-0003-4153-4105} 
\and
Kaneesha Moore\inst{1}\orcidID{0009-0000-1197-3228} 
\and
Sudip Mittal\inst{1}\orcidID{0000-0001-9151-8347} 
\and
Edward Zieglar\inst{3}\orcidID{0000-0001-5107-2125}}
\authorrunning{M. Duclos et al.}
%
\institute{Mississippi State University, Mississippi State, MS 39762, USA
\email{\{md128,iaf28,kkm267\}@msstate.edu,mittal@cse.msstate.edu}\and
National Security Agency, Fort George G. Meade, MD 20755, USA
\email{evziegl@nsa.gov}\\}
\maketitle              
\begin{abstract}
Formal methods can be used for verifying security protocols, but their adoption can be hindered by the complexity of translating natural language protocol specifications into formal representations. In this paper, we introduce ModelForge, a novel tool that automates the translation of protocol specifications for the Cryptographic Protocol Shapes Analyzer (CPSA). By leveraging advances in Natural Language Processing (NLP) and Generative AI (GenAI), ModelForge processes protocol specifications and generates a CPSA protocol definition. This approach reduces the manual effort required, making formal analysis more accessible. We evaluate ModelForge by fine-tuning a large language model (LLM) to generate protocol definitions for CPSA, comparing its performance with other popular LLMs. The results from our evaluation show that ModelForge consistently produces quality outputs, excelling in syntactic accuracy, though some refinement is needed to handle certain protocol details. The contributions of this work include the architecture and proof of concept for a translating tool designed to simplify the adoption of formal methods in the development of security protocols.

\end{abstract}

\section{Introduction}
\label{sec:intro}
The current standards process used by the Internet Engineering Task Force (IETF) for reviewing security protocols is hindered by a lack of verification of the stated protocol properties. This shortcoming primarily arises from the absence of formal proofs, which are essential for thorough verification. In an effort to enhance the standards process, the IETF is encouraging protocol developers to incorporate formal methods analysis and validation into their work \cite{ietf2023ufmrg}. Formal methods are mathematical techniques used to prove the correctness and security properties of protocols. However, formal methods analysis is a complex area that demands specialized expertise, rendering it inaccessible to some security protocol developers \cite{gritzalis1999security}. This paper aims to support the IETF’s efforts by simplifying formal methods analysis and validation, thereby making these tasks more accessible to security protocol developers.

One approach to formal methods analysis, as described by Meadows \cite{meadows1992applying} and originally suggested by Kemmerer \cite{kemmerer1989using}, involves modeling a protocol in a formal language. A verification system is then employed to validate these models against the stated protocol properties (e.g., secrecy and authentication). Protocol developers perform analysis and verification using tools such as Cryptographic Protocol Shapes Analyzer (CPSA), Maude-NPA, Tamarin, and ProVerif. However, this task is both tedious and labor-intensive, often requiring a \textit{domain expert} to parse and convert protocol specifications—typically expressed in natural language \cite{rfc1334,rfc7296,rfc8446}—into models or protocol definitions compatible with a formal methods tool \cite{ietf2023ufmrg}. Consequently, accurately translating protocol specifications into a usable form for formal methods tools poses a significant challenge \cite{gritzalis1999security}.

\begin{figure}
    \centering
    \includegraphics[width=.7\linewidth]{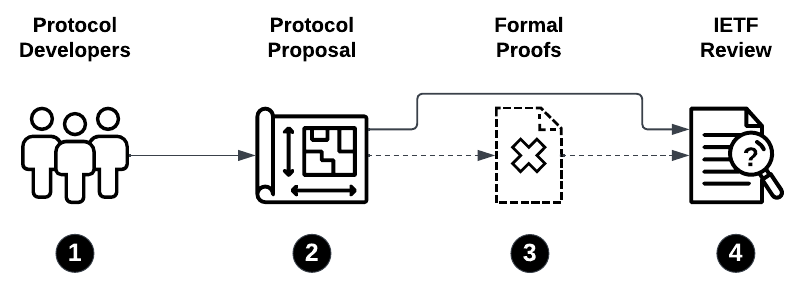}
\caption{Overview of the IETF standards process for security protocols, highlighting the lack of formal verification at key stages. ModelForge aims to automate aspects of formal methods analysis to address these gaps. (1) Developers design a new protocol, (2) compile protocol specifications into a proposal, (3) often submitting it without formal proofs, (4) leaving the IETF to establish formal proofs during review.}    \label{fig:ietf_process}
\end{figure}

As illustrated in Figure \ref{fig:ietf_process}, the current IETF standards process could be improved by making formal methods analysis tools more accessible to protocol developers. Increased accessibility would likely encourage security protocol developers to integrate these tools into their workflow, enabling them to perform protocol analysis during the specification development phase, even before submitting a protocol proposal to the IETF. We envision an enhanced workflow where a translator tool is capable of accepting protocol specifications and turning them into protocol definitions or models for a formal methods tool. Such a system could potentially automate a portion of the creation of security protocol definitions, reducing the manual effort required and improving efficiency. Subsequently, this translator tool would align with one of the goals of the Usable Formal Methods Research Group (UFMRG), established by the IETF in January 2023 \cite{ietf2023ufmrg}, which seeks to understand how formal methods can be incorporated into the development of specifications for security protocols.

Building on our objective of making formal methods analysis more accessible, we propose ModelForge, a translating tool that leverages recent advances in Natural Language Processing (NLP) and Generative AI (GenAI). Our tool feeds protocol specifications into a fine-tuned Large Language Model (LLM), which then automatically generates candidate protocol definitions or models. ModelForge is specifically designed with CPSA as the target protocol analyzer, requiring the LLM to output the protocol definitions using CPSA syntax. The CPSA formal methods tool necessitates a protocol definition (model) and a partial description of an execution (protocol skeleton) as inputs, both formatted as structured text using symbolic expressions (s-expressions) \cite{cpsa41manual2023}. The structured output candidates generated by ModelForge will need to be reviewed and revised by a domain expert. This step is to ensure that protocol definitions accurately represent the protocol specifications before undergoing analysis. This novel approach has the potential to enhance the overall efficiency of the secure protocol formal analysis process by reducing the manual effort and complexity associated with translating specifications into the format required by CPSA, thereby making formal analysis more accessible.

Our key contributions are: 1) the development of an architecture for translating natural language protocol specifications into CPSA protocol definitions, 2) the validation of this architecture through a proof-of-concept implementation, and 3) the evaluation of various LLMs in translating protocol specifications into CPSA protocol definitions. In summary, our paper presents a methodology that lowers the barriers to adopting formal methods tools in the field of security protocol development and analysis.

To address the complexity and inaccessibility of formal methods in security protocol development, as we discussed in the introduction, we now focus on key concepts and related work in formal methods, NLP, and Generative AI, which provide the basis for our proposed solution, ModelForge.

\section{Background and Related Work}
\label{sec:background}
Despite advancements in Natural Language Processing (NLP), translating protocol specifications into formal definitions remains a challenge across various tools, including CPSA. Automating this translation process offers a potential solution. This section reviews related work in formal methods, NLP, and large language models (LLMs) as a foundation for addressing this challenge.

\subsection{Internet Engineering Task Force}
\label{sec:ietf}
The Internet Engineering Task Force (IETF) develops standards and security protocols through the Request for Comments (RFC) process, which relies heavily on natural language to promote discussion and consensus. In January 2023, the IETF established the Usable Formal Methods Research Group (UFMRG) to promote the use of formal methods in protocol standards development \cite{ietf2023ufmrg}.

\subsection{Formal Methods in Protocol Analysis}
\label{sec:formal_methods}
Protocol developers and researchers can utilize different formal methods tools to verify the security goals of a protocol. In the context of such tools, formal methods are expressive mathematical models used to prove theorems constructed from a protocol's stated properties. These methods fall under domains such as model checking and logical inference. As defined by Lal et. al, the emphasis within model checking lies in exploring all possible states and transitions in a systematic nature, using abstractions to optimize the verification analysis. Whereas, logical inference, the domain in which our work resides, uses mathematical reasoning of the system to determine the validity of the protocol analysis \cite{Lal2011ApproachesTF}.

\subsection{Cryptographic Protocol Shapes Analyzer}

The MITRE Corporation developed the Cryptographic Protocol Shapes Analyzer (CPSA) \cite{cpsa41manual2023}, an open-source tool for analyzing cryptographic protocols using logical inference within a strand-space model. CPSA acts as a model finder, exploring protocol behaviors through roles, messages, variables, and their assumptions, defined using a LISP (LISt Processing)-like syntax.”

CPSA uses partial executions, known as skeletons, which instantiate strands and roles with variable assumptions to identify possible protocol behaviors. Figure~\ref{fig:cpsa_model} shows an example from the CPSA manual. In a Dolev-Yao network \cite{Dolev1981OnTS}, where an adversary can intercept, alter, and fabricate messages, CPSA finds all distinct protocol executions and outputs them as shapes \cite{liskov2011completeness}.

Shapes represent distinct protocol executions, and each contains analysis sentences that capture “all that can be learned from a CPSA run” \cite{ramsdell2015cpsa}, enabling derivation of security goals and proofs. We selected CPSA for our work due to our familiarity with the tool and access to experts.

\begin{figure}[htbp]
\begin{tcolorbox}[
colback=cyan!4!white!95!blue,
colframe=gray!90!black,
colbacktitle=gray!80!black, 
left=0.5mm, 
right=0.5mm, 
boxrule=0.50pt]
\scriptsize
{\fontfamily{qcr}\selectfont
\begin{Verbatim}
(defprotocol blanchet basic
  (defrole init
    (vars (a b akey) (s skey) (d data))
    (trace (send (enc (enc s (invk a)) b))
           (recv (enc d s)))
    (uniq-orig s))
  (defrole resp
    (vars (a b akey) (s skey) (d data))
    (trace (recv (enc (enc s (invk a)) b))
           (send (enc d s)))
    (uniq-orig d)))

(defskeleton blanchet
  (vars (a b akey) (s skey) (d data))
  (defstrand init 2 (a a) (b b) (s s) (d d))
  (deflistener d)
  (non-orig (invk b)))
  
(defskeleton blanchet
  (vars (a b akey) (s skey) (d data))
  (defstrand resp 2 (a a) (b b) (s s) (d d))
  (deflistener d)
  (non-orig (invk a) (invk b)))
    
\end{Verbatim}
}
\end{tcolorbox}
\caption{Blanchet’s Protocol, defined in CPSA syntax, consists of two steps. First, the initiator (init) sends a symmetric key s to the responder (resp), signed with init’s private key (a) and encrypted with resp’s public key (b). Then, resp sends data d back to init, encrypted with s.}
\label{fig:cpsa_model}
\end{figure}

\subsection{Large Language Models}

LLMs are versatile models trained on large datasets to perform a wide range of tasks across various domains. Their development goes back to the rise of foundational models in natural language processing (NLP) \cite{bommasani2021opportunities}. Foundational models are designed to be highly adaptable, using diverse training data to enable fine-tuning for multiple tasks. However, their broad training scope can lead to limitations in accuracy when applied to specialized fields.

\subsection{ChemLLM}
Foundational LLMs like OpenAI’s ChatGPT and Meta’s Llama are widely used, but specialized domains, such as chemistry, benefit from domain-specific models. Zhang et al. \cite{zhang2024chemllm} emphasize that foundational models often lack chemical knowledge and propose ChemLLM, a chemistry-specific LLM fine-tuned from a foundational model. They also demonstrate the use of synthetic data to train the model when organic data is limited.

\subsection{StrucBench}
Extending LLMs beyond simple text generation can be challenging. Tang et al. \cite{tang2023struc} note that LLMs, such as GPT-4, struggle with generating intricate formatted text including complex structured data (e.g., tables). These limitations are likely due to the way LLMs are trained to mimic patterns of human language and the token requirements for structured output. Tang et al. propose the  Format Chain-of-Thought (FormatCOT) methodology for generate fine-tuning pairs via self-instruction. To address the research gaps with LLMs used for structured output including systemic analyses, they also introduce the STRUC-BENCH benchmark that focuses on structured text generation (e.g., HTML, LaTeX).

\section{ModelForge Architecture}
\begin{figure}[ht]
\centering
\includegraphics[width=1\linewidth]{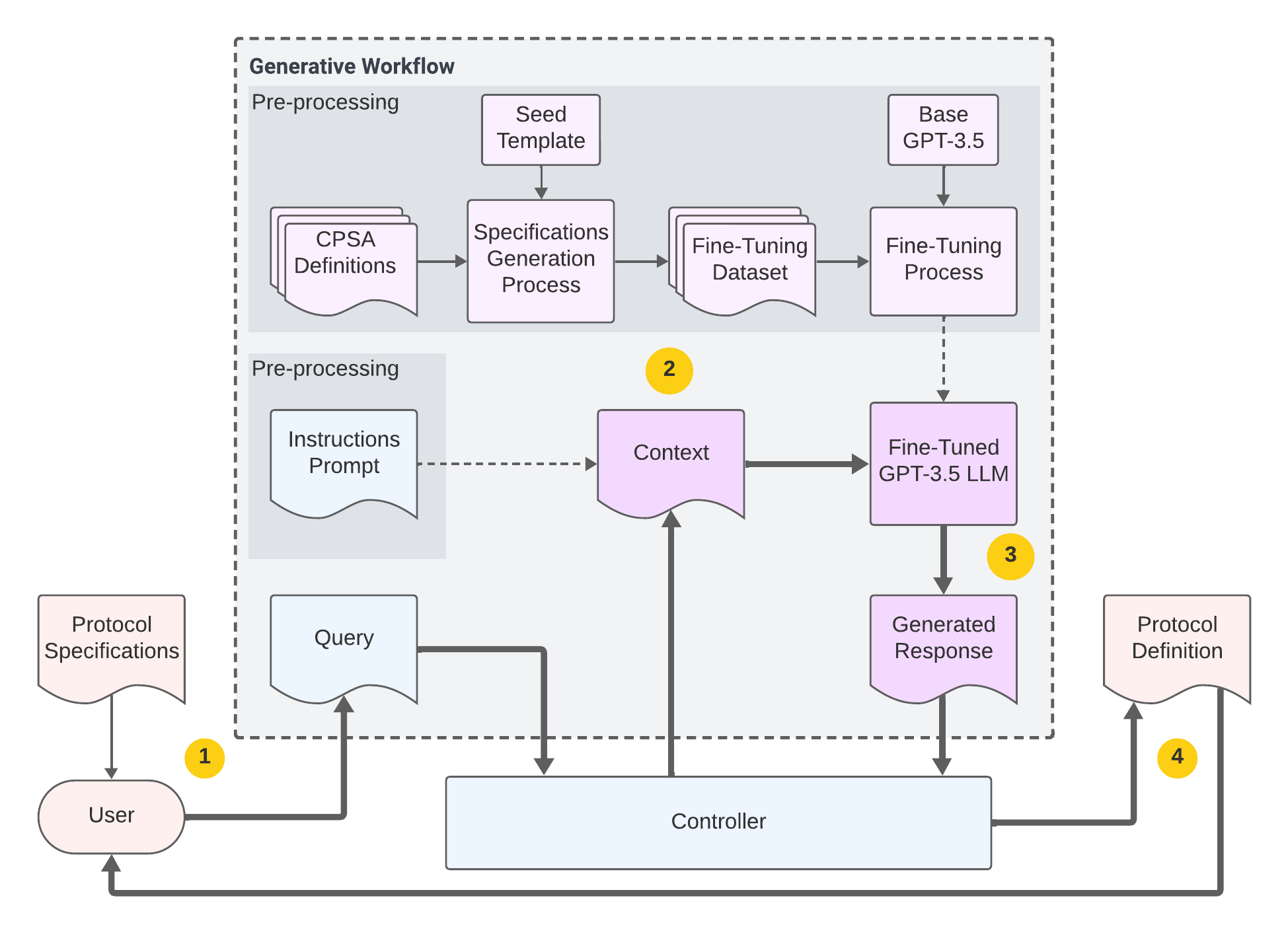}
\caption{ModelForge architecture: (1) A user submits protocol specifications as a query. (2) The controller combines the query with an instruction prompt and sends it to the fine-tuned LLM. (3) The LLM generates a CPSA protocol definition. (4) The output is then delivered to the user for validation.}
\label{fig:architecture}
\end{figure}

In this section, we present the modular architecture of ModelForge, a translator designed to convert natural language secure protocol specifications into CPSA protocol definitions. Figure~\ref{fig:architecture} illustrates how the architecture centers around a fine-tuned LLM for generating protocol definitions in response to user defined queries. Like CPSA, ModelForge is designed for researchers and developers working with cryptographic protocols and formal methods. The main components and data flow are illustrated in Figure~\ref{fig:architecture}.

\subsection{Query}
\label{sec:query}
A user-provided query serves as both the input and the starting point for ModelForge. The query consists of a natural language protocol specifications. Figure~\ref{fig:user_query} presents an example of a simple user query.

\begin{figure}[ht]
\begin{tcolorbox}[
colback=cyan!4!white!95!blue,
colframe=gray!90!black,
colbacktitle=gray!80!black, 
left=0.5mm, 
right=0.5mm, 
boxrule=0.50pt]
\scriptsize
\raggedright
{\fontfamily{qcr}\selectfont
\textbf{Protocol Description:}
Alice and Bob require a secure protocol for mutual authentication over an insecure network. Using the specifications below, create a CPSA protocol definition. The protocol must employ public-key encryption[...]
\vspace{\baselineskip}

\textbf{Initiation Sequence:}
Alice sends a message to Bob including Alice's identity (a) and her nonce (n1), encrypted with Bob's public key. Upon receiving Alice's message, Bob sends to Alice the nonce (n1) received from Alice, his identity (b), and his nonce (n2), encrypted with Alice's public key. Alice decrypts Bob's message then sends Bob's nonce (n2) back to him.
}
\end{tcolorbox}
    \caption{Example of a user-submitted query for a secure protocol using public-key encryption. This query serves as input for ModelForge.}
    \label{fig:user_query}
\end{figure}

\subsection{Generative Workflow}
\label{sec:phase2}
The keystone of the generative workflow in our architecture is a fine-tuned LLM, a subset of generative AI focused specifically on text generation. In our use case, this model generates a CPSA protocol definition from natural language specifications, aiding in the formal methods analysis of security protocols.

\subsection{Pre-processing}
Before the generative workflow can be utilized, several components must be initialized as part of a pre-processing sequence. These components, highlighted in a darker shade in Figure~\ref{fig:architecture}, consist of pre-processing tasks, loosely categorized into dataset-related, fine-tuning, and configuration tasks.

\subsection{Fine-Tuning Dataset}
One of the initial challenges we faced was the scarcity of CPSA protocol definitions available to us for our fine-tuning goal. To address this issue, we utilized GPT-4, to generate ten seed templates of CPSA protocol definitions. These templates served as the question portion of a question-and-answer (Q\&A) pair. GPT-4 was further employed to generate corresponding protocol specifications, creating synthetic Q\&A pairs of CPSA protocol definitions and their associated specifications. 
After forming these pairs, GPT-4 was leveraged once again to introduce variance into the existing set of Q\&A pairs, thereby increasing the number of pairs available for the fine-tuning process.

\subsection{Fine-Tuning Process}
With a curated dataset of synthetic Q\&A pairs in place, the next step involved fine-tuning a pre-trained LLM to ensure it could accurately generate protocol definitions, a task requiring domain-specific adjustments. Fine-tuning is a method used to update the weights of a pre-trained model by further training it on specific data. This process is referred to as supervised fine-tuning when it uses labeled data tailored to a particular task \cite{brown2020language}. In other words, fine-tuning enables us to modify the LLM in two ways: 1) inject it with additional knowledge, and 2) altering its behavior to perform specific tasks more effectively. In Section~\ref{sec:setup}, we detail the fine-tuning process associated with OpenAI’s GPT-3.5 model, the LLM selected for ModelForge.

\subsection{Configuration}
\label{sec:config}
The configuration components are the settings and parameters that guide the behavior of the LLM during the generation process. In our architecture, this consists of a single but important configuration component: a pre-defined instructions prompt. This prompt includes specialized instructions and a placeholder for the user query.

\subsection{Context}
The context is obtained by concatenating two components: 1) the user query and 2) the instructions prompt. These two components form the context, and is later fed by the controller as an input into the LLM for processing.

\subsection{Fine-Tuned LLM}
\label{sec:finetuned_llm}
Core to our architecture, is an LLM responsible for translating the input into an output that contains a CPSA protocol definition. To accomplish this task, the LLM relies on three types of knowledge: 1) knowledge acquired during an initial pre-training phase, 2) knowledge gained through the fine-tuning process, and 3) context-specific knowledge provided by the user query. In Section~\ref{sec:setup}, we detail the specifics of using GPT-3.5-turbo from OpenAI, the LLM we selected for the implementation of our translator.

\subsection{Generated Response}
\label{sec:generated_response}
The output out of the LLM consists of structured text, made of symbolic expressions that form a CPSA protocol definition. See Figure \ref{fig:llm_response} for an output example.

\begin{figure}
\begin{tcolorbox}[
colback=cyan!4!white!95!blue,
colframe=gray!90!black,
colbacktitle=gray!80!black, 
left=0.5mm, 
right=0.5mm, 
boxrule=0.50pt]
\scriptsize
{\fontfamily{qcr}\selectfont
\begin{Verbatim}
(herald "Mutual Authentication Protocol 
 with Public Key Encryption")

(defprotocol mutual-auth basic
   (defrole init
     (vars (a name) (b name) (n1 text) (n2 text) 
     (pubk data))
     (trace
       (send (enc (cat a n1) (pubk b)))
       (recv (enc (cat n1 n2) (pubk a)))
       (send n2)))
\end{Verbatim}
}
\end{tcolorbox}
    \caption{Example of a generated CPSA protocol definition by ModelForge, illustrating the output format using s-expressions for protocol roles and trace sequences.}
    \label{fig:llm_response}

\end{figure}
\vspace{-6mm}

\subsection{Controller}
\label{sec:controller}
The controller’s role is to orchestrate the sequence of execution between the architectural components. 
Additionally, the controller is responsible for initializing the system and setting up the necessary parameters and states.

\subsection{Data Flow}
\label{sec:data_flow}
As illustrated in Figure~\ref{fig:architecture}, the flow of data through the main components of the architecture is described below.

\noindent(1) User Query: The process begins when a user submits a query containing secure protocol specifications, similar to the example in Figure~\ref{fig:user_query}. 

\noindent(2) Query and Prompt Concatenation: The next step consists of concatenating the user query with instructions prompt. This will form the input for the LLM.

\noindent(3) LLM Response Generation: The combined input is then fed into a fine-tuned LLM, which is instructed to generate a CPSA protocol definition based on the secure protocol specifications contained within the input.

\noindent(4) Output Delivery: Finally, the controller receives the response generated by the LLM and directs it back to the user.

\section{Experimental Results}
\label{sec:exp_results}
In this section, we present the findings from our experiments, including both a quantitative and qualitative evaluation methods of our fine-tuned translator model. We also provide information on the technical setup of our experiment, along with details on the survey instrument used to collect data for the evaluation. Finally, we present and discuss the results, offering insights into the effectiveness of our approach.

\subsection{Evaluation}
\label{sec:eval}
To assess the effectiveness of ModelForge in generating CPSA protocol definitions, we conducted an evaluation using both quantitative and qualitative methods. The evaluation involved 4 participants with relevant experience ranging from less than one year to 10 or more years. These participants reviewed the outputs generated by various LLMs, rating their correctness, clarity, and completeness, while also providing feedback on the outputs' strengths and weaknesses. This allowed us to gather valuable data on the LLMs’ effectiveness in meeting the requirements of CPSA protocol definitions. 
The data collection was carried through a survey implemented using Google Forms. The survey presented participants with three distinct queries and 
a total of nine corresponding CPSA protocol definitions. Table~\ref{tab:survey_criteria} provides the specific criteria and definitions used in the survey.

\begin{table}[htbp]
    \centering
    \caption{Evaluation criteria for assessing CPSA protocol definitions, including correctness, clarity, and completeness for formal analysis.}
    \begin{tabularx}{\linewidth}{p{0.35\linewidth} X}
        \hline
        \textbf{Criterion} & \textbf{Description} \\
        \hline
        Correctness & Output uses proper CPSA syntax. \\
        Clarity & Output is easy to read and well-structured. \\
        Completeness & Output covers the main aspects of the query. \\
        \hline
    \end{tabularx}
    \label{tab:survey_criteria}
\end{table}

\subsection{Quantitative Method}

Our quantitative portion of the evaluation, focused on assessing how well ModelForge, along with other popular LLMs (see Table \ref{tab:survey_llms}), could generate an output that aligns with the syntactical requirements of CPSA, readability of protocol definitions, and the intent of the user query.

To assess the protocol definitions, we collected ratings from subject-matter experts who evaluated how closely each model’s output approximated a complete, accurate and easily readable CPSA protocol definition. 

For each output, experts were asked to rate its correctness, clarity, and completeness on a scale from 1 to 5, with 1 indicating an unusable output and 5 indicating a very good output. Table~\ref{tab:survey_criteria} provides the criteria used in the survey.

The experts’ ratings provided data on each LLM’s performance, highlighting their ability or inability in generating CPSA protocol definitions. A summary of the ratings provided by the experts is presented in Table~ \ref{tab:survey_results}.

\subsection{Qualitative Method}
For the qualitative portion, we again asked the same experts to provide feedback on the strengths and weaknesses of the generated CPSA protocol definitions. 
Experts were asked to describe the strengths and weaknesses of each generated protocol definition, providing detailed feedback on aspects that are critical for the practical application of CPSA protocols but difficult to quantify. We present a summary of the qualitative results in Table \ref{tab:qual_modelforge}.

\subsection{Experimental Setup}
\label{sec:setup}
To gather experimental results for ModelForge, we implemented a proof of concept based on the architecture depicted in Figure~\ref{fig:architecture}. This setup enabled us to evaluate the performance of ModelForge by having human experts rate the outputs it generated. Below, we describe the technical details of our implementation.

\noindent \textbf{Framework:}
At the core of our implementation is LlamaIndex, a modern data framework for LLM-based applications, built using Python \cite{rau2023context,rau2024retrieval,braunschweiler2023evaluating,Liu_LlamaIndex_2022}. We chose LlamaIndex due to our familiarity with the framework and its ease of use in working with LLMs.

\noindent \textbf{LLM and Fine-Tuning:}
Building on the LlamaIndex framework, we used a fine-tuned version of OpenAI’s GPT-3.5-turbo model to generate CPSA protocol definitions from natural language specifications. The base model, \textit{gpt-3.5-turbo}, was fine-tuned to handle the structured nature of CPSA syntax.

Our fine-tuning process utilized a carefully curated dataset of 340 synthetic Q\&A pairs, partitioned into a 70\% training set and a 30\% testing set. The model underwent 3 epochs of training, processing a total of 568,200 tokens. This resulted in a training loss of 0.2976 and a validation loss of 0.3902, indicating effective generalization of the task of generating CPSA protocol definitions from input specifications.

For the generation phases of both the fine-tuning dataset and protocol definitions, we intentionally set the temperature parameter to 0. This parameter controls the randomness of the LLM's outputs. This choice of temperature encourages more deterministic results, prioritizing consistency and adherence to the protocol specifications over creative variations. By minimizing randomness, we aimed to ensure that ModelForge's outputs aligned with the structured nature of CPSA syntax (see Section~\ref{fig:cpsa_model}) and the specific requirements of each protocol definition.

\subsection{Results}
\label{sec:results}
Using our experimental setup, we evaluated the performance of ModelForge as well as other LLMs (see Table~\ref{tab:survey_llms}) for comparison. In Tables~\ref{tab:survey_results} and~\ref{tab:qual_modelforge} we present the results for both quantitative and qualitative methods of our evaluation.

\begin{table}[ht]
    \centering
    \caption{List of additional LLMs evaluated alongside ModelForge.}
    \begin{tabular}{p{0.15\linewidth} p{0.15\linewidth} p{0.15\linewidth}p{0.50\linewidth}}
        \hline
        \textbf{LLM} & \textbf{Author} & \textbf Parameters& \textbf{Description}\\
        \hline
        Claude 3.5 & Anthropic & Unknown & Conversational model for dialogue and interactive tasks. \\
        LLama 3 & Meta & 70B &Multilingual model for a wide variety of tasks. \\
        GPT-4o & OpenAI & $>$175B &Versatile model capable of a wide variety of tasks. \\
        Codestral & Mistral AI &22B & Specialized in generating and understanding code. \\
        \hline
    \end{tabular} 
    \label{tab:survey_llms}
\end{table}

The results of our quantitative study found in Table~\ref{tab:survey_results}, summarizes the experts’ survey ratings across the three queries and nine models evaluated. The data reflects the experts’ assessments of each model’s correctness, clarity, completeness, as defined in Table~\ref{tab:survey_criteria}. These ratings provide valuable insights into the relative effectiveness of ModelForge, compared to other leading LLMs.

\begin{table*}[ht]
    \centering
    \caption{Normalized average ratings provided by experts for each query and model. Metrics include correctness, clarity, and completeness for CPSA protocol definitions.}
    \begin{tabularx}{\linewidth}{p{0.15\linewidth} p{0.2\linewidth} >{\centering\arraybackslash}X >{\centering\arraybackslash}X >{\centering\arraybackslash}X}
        \hline
        \textbf{Query} & \textbf{Model} & \textbf{Correctness} & \textbf{Clarity} & \textbf{Completeness} \\
        \hline
        \multirow{5}{*}{Query 1} & \LLMOne   & 0.00 & 0.50 & 0.33 \\
                                 & \LLMTwo   & 0.50 & 0.50 & 0.33 \\
                                 & \LLMThree & 0.25 & 0.50 & 0.42 \\
                                 & \LLMFour  & 0.00 & 0.00 & 0.00 \\
                                 & \LLMFive  & 0.67 & 0.75 & 0.67 \\
        \hline
        \multirow{5}{*}{Query 2} & \LLMOne   & 0.00 & 0.17 & 0.08 \\
                                 & \LLMTwo   & 0.75 & 0.92 & 0.83 \\
                                 & \LLMThree & 0.00 & 0.00 & 0.00 \\
                                 & \LLMFour  & 0.00 & 0.17 & 0.25 \\
                                 & \LLMFive  & 0.83 & 0.92 & 0.92 \\
        \hline
        \multirow{5}{*}{Query 3} & \LLMOne   & 0.00 & 0.33 & 0.25 \\
                                 & \LLMTwo   & 0.50 & 0.50 & 0.50 \\
                                 & \LLMThree & 0.80 & 0.80 & 0.80 \\
                                 & \LLMFour  & 0.00 & 0.00 & 0.33 \\
                                 & \LLMFive  & 0.75 & 0.50 & 0.50 \\
        \hline
    \end{tabularx}
    \label{tab:survey_results}
\end{table*}
ModelForge, as indicated by the results, demonstrates consistent performance across correctness, clarity, and completeness, particularly excelling with Query 2, where it achieved some of the highest ratings across all models. However, there is room for improvement with Query 3, where GPT-4o outperformed ModelForge. Despite this, ModelForge remains competitive, consistently ranking among the top-performers overall.

From a qualitative perspective, the feedback on the strengths and weaknesses provided by survey participants was also insightful. As summarized in Table~\ref{tab:qual_modelforge}, ModelForge was regarded as having the best instance of CPSA syntax among the evaluated models, but it still exhibited specific weaknesses, particularly in protocol handling, such as issues with the use of public keys. While the structure and syntax of its outputs were praised, the feedback highlighted opportunities for further refinement to achieve greater CPSA compatibility.
\begin{table*}[ht]
    \centering
    \caption{Summary of qualitative feedback on ModelForge’s performance, highlighting strengths in CPSA syntax generation and areas requiring improvement.}
    \begin{tabularx}{\textwidth}{p{0.25\linewidth} X}
        \hline
        \textbf{Aspect} & \textbf{Feedback} \\
        \hline
        Strengths & Best CPSA syntax so far. Strong structure and formatting, recognized as clear. \\
        \hline
        Weaknesses & Public key issue: Sending out the public key, which was not part of the protocol definition. Limited CPSA compliance, needs refinement for full compatibility. \\
        \hline
    \end{tabularx}
    \label{tab:qual_modelforge}
\end{table*}

\section{Limitations and Future Work}
\label{sec:future_work}
While our evaluation in Section~\ref{sec:eval} demonstrates that ModelForge can generate syntactically correct CPSA definitions, particularly excelling in certain queries, there are notable limitations. In this section, we explore some of these limitations and propose avenues for future improvement.

ModelForge’s initial outputs generally adhered to the CPSA syntax format, however, they often required corrections. As such, human experts should remain an integral part of any workflow. More precisely, a domain expert should carefully review the AI-generated CPSA protocol definitions to ensure they are correct, complete, and secure. A Human-in-the-Loop validation is an efficient way to increase automation while minimizing risks caused by limitations.  

One notable limitation of ModelForge is that it performs better when variable names are provided as part of the user query. This presents a challenge when translating documents such as RFCs, where variable names are not always explicitly stated. To address this challenge, future work could explore methods such as reinforcement learning from human feedback to establish a basis for automatically inferring or generating variable names.

Another significant challenge arises when working with RFCs, which often contain a considerable amount of unnecessary content to form a protocol definition. One potential solution to this issue could involve training a separate AI model designed to parse and extract only the necessary content for translation into a protocol definition. 

Addressing these limitations, such as inferring variable names and content extraction from RFCs, will be key milestones in enhancing ModelForge’s capabilities. Future research could explore larger-scale evaluations and the development of additional models to handle more complex protocol features. In conclusion, our work demonstrates the potential of AI-driven tools in simplifying formal methods analysis, moving us closer to the IETF’s goal of integrating formal methods into security protocol development.

\section{Conclusion}
In this paper, we presented ModelForge, a novel translating tool designed to facilitate the integration of formal methods analysis into the development of security protocols. Our evaluation, detailed in Section~\ref{sec:eval}, demonstrates ModelForge's strong performance. It consistently performs well and, in several instances, even outperforms more recent foundational LLMs. It is important to note that only ModelForge had been fine-tuned for the specific task, further highlighting the effectiveness of domain-specific language models.

Looking ahead, future research could involve larger-scale evaluations across a broader range of protocol specifications, as well as the development of an ensemble of models to handle more complex features within CPSA, such as Diffie-Hellman algebra, goals, and listeners.

Overall, the contributions of our work represent a step towards making formal methods more accessible, supporting the Internet Engineering Task Force’s (IETF) efforts to improve the verification of security protocol standards.

\begin{credits}
\subsubsection{\ackname}
This work was supported by the PATENT Lab (Predictive Analytics and Technology Integration Laboratory) in the Department of Computer Science and Engineering at Mississippi State University. The authors would also like to thank the participants who completed the survey. We extend our sincere gratitude to Enis Golaszewski for his invaluable guidance in utilizing CPSA, which significantly contributed to this research.
\subsubsection{\discintname}

The authors have no competing interests to declare that are relevant to the content of this article. 

\end{credits}

%
%
%

\end{document}